\documentclass[prb,preprint]{revtex4}

\usepackage{graphicx}
\usepackage{bm}
\usepackage{subfigure}
\usepackage{epstopdf}
\usepackage{hyperref}

\begin{document}

\title{Self-localization of a single hole in Mott antiferromagnets}

\author{Zheng Zhu$^1$, Hong-Chen Jiang$^{2,3},$ Yang Qi$^1$, Chu-Shun Tian$^1$, and Zheng-Yu Weng$^1$}
\affiliation{$^{1}$Institute for Advanced Study, Tsinghua University, Beijing, 100084, P.
R. China\\
$^2$Kavli Institute for Theoretical Physics, University of California, Santa
Barbara, CA, 93106-4030, U.S.A.\\
$^3$Center for Quantum Information, IIIS, Tsinghua University, Beijing, 100084, China}
\date{\today}
\begin{abstract}
A long-standing
issue\cite{Anderson,Anderson90PRL,Shraiman1988,SCBA1,SCBA2,SCBA3,SCBA4,Dagotto1994,ED,Shih97,White2,
Laughlin97,Weng1996,Weng1997,Weng2001} in the physics of strongly
correlated electronic systems is whether the motion of a single hole
in quantum antiferromagnets can be understood in terms of the
quasiparticle picture. Very recently, investigations of this issue
have been within the experimental reach\cite{STM}. Here we perform a
large-scale density matrix renormalization group study, and provide
the first unambiguous numerical evidence showing that in ladder
systems, a single hole doped in the Mott antiferromagnet does not
behave as a quasiparticle. Specifically, the injected hole is found
to be always localized as long as the leg number is larger than one,
with a vanishing quasiparticle weight and a localization length
monotonically decreasing with the leg number. In addition, the
single hole self-localization is insensitive to the parity
(even-odd) of the leg number. Our findings may advance conceptual
developments in different fields of condensed matter physics. First
of all, the intriguing self-localization phenomenon is of pure
strong correlation origin free of extrinsic disorders. Therefore, it
is in sharp contrast to the well-known Anderson
localization\cite{Anderson58} and recently found many-body
localization\cite{Basko06}, where extrinsic disordered potentials
play crucial roles. Second, they confirm the analytical
predictions\cite{Weng1997,Weng2001} of the so-called phase string
theory\cite{Weng1996,Weng1997}, suggesting that the phase string
effect lies in the core of the physics of doped Mott
antiferromagnets.
\end{abstract}

\maketitle

The doped Mott insulator is generally believed to be the
prototypical model of many realistic strongly correlated systems
notably high $T_c$ cuprates\cite{Anderson,anderson87,Lee06}, and is
a fundamental problem of great challenge in condensed matter
physics. Important insights into doped Mott insulators may be gained
by inspecting a special case namely the single hole doped Mott
insulator. This has been under intense
investigations\cite{Anderson,Anderson90PRL,Shraiman1988,SCBA1,SCBA2,SCBA3,SCBA4,
Dagotto1994,ED,Shih97,White2,Laughlin97,Weng1996,Weng1997,Weng2001}
for over decades, but remained highly controversial. On the one
hand, a quasiparticle weight was found by using either analytic
approaches (in combination with the self-consistent Born
approximation)\cite{SCBA1,SCBA2,SCBA3,SCBA4} or finite-size exact
diagonalization\cite{Dagotto1994} on lattices up to $32$
sites\cite{ED}. This indicates that an injected hole with a Fermi
momentum maintains coherent motion in the quantum antiferromagnetic
spin background. On the other hand, the validity of such a
quasiparticle picture has been seriously
questioned\cite{Anderson90PRL,Laughlin97,Weng1996,Weng1997,Weng2001}
by various authors. It was first argued by
Anderson\cite{Anderson90PRL} that the quasiparticle weight vanishes
when a hole is injected into the Mott insulator. Later, this crucial
observation received justification on a completely microscopic
level: by identifying the so-called phase string
effect\cite{Weng1996,Weng1997}, it was proved that the quasiparticle
weight indeed vanishes at the ground states. In addition, it was
further predicted\cite{Weng2001} that the non-perturbative phase
string effect may lead to self-localization of the injected hole.

On the experimental side, in earlier angle-resolved photoemission
spectroscopy (ARPES) studies\cite{Shen95,ARPES,Shen04}, broad
spectral features have been observed in materials such as ${\rm
Ca}_2{\rm CuO}_2{\rm Cl}_2$ and ${\rm Sr}_2{\rm CuO}_2{\rm Cl}_2$.
This shows indirectly that the motion of a single hole in quantum
antiferromagnets may not be understood in terms of the quasiparticle
picture\cite{Weng2001,Shen04}. Very recently, an unprecedented
degree of control has been reached in experiments with the ${\rm
Ca}_2{\rm CuO}_2{\rm Cl}_2$ parent Mott insulators, opening a new
route for experimental study of a single charge doped Mott
insulator. In particular, scanning tunneling microscope (STM)
experiments on the atomic scale electronic structure of this
material have unveiled a striking phenomenon: a single electron
donated by surface defect creates an electronic state strongly
localized in space\cite{STM}. Consistent with the earlier ARPES
experiments above, the observed localization of the single doped
charge also does not support the quasiparticle picture.

While numerical simulation serves as a powerful tool for resolving
this puzzle, a crucial point is how to extend results obtained from
finite-size numerical simulations to the large sample limit. To the
best of our knowledge, this issue has not yet been addressed in the
literature. The purpose of this work is to carry out the first
large-scale numerical study and resolve the important issue of the
validity of the quasiparticle picture in lightly doped Mott
antiferromagnets.

To this end, we adopt the unbiased density matrix renormalization
group (DMRG) method\cite{White1992} to extensively investigate this
problem in ladder systems. We find that if a ladder sample is
sufficiently long, the injected hole is always localized in the
quantum antiferromagnetic spin background, once the leg number is
larger than one. In fact, the localization length gets continuously
reduced with the increase of the leg number, pointing to a strong
localization in the two-dimensional limit. Contrary to this, if the
sample is sufficiently short, the injected hole remains itinerant.
This suggests that quasiparticle behavior found in earlier numerical
study\cite{Dagotto1994,ED} is likely a small-size effect. The
intriguing phenomenon of self-localization of a single hole in Mott
antiferromagnets and the vanishing quasiparticle spectral weight in
the large sample limit constitute the unambiguous numerical evidence
showing that a single hole doped into the Mott antiferromagnet no
longer behaves like a quasiparticle. These support the analytic
predictions\cite{Weng1997,Weng2001} of the phase string theory for
doped quantum antiferromagnets.

{\it Model and numerical method.---} The DMRG calculation is
performed for the $t$-$J$ model Hamiltonian,
\begin{equation}
H_{tJ}=-t\sum_{\langle {ij}\rangle \sigma}({c_{i\sigma}^{\dag
}c_{j\sigma }+h.c.}) + J\sum_{\langle {ij}\rangle }(\bm S_i\cdot
\bm S_j-\frac{1}{4}n_{i}n_{j}). \label{a}
\end{equation}%
Here, ${{c_{i\sigma }^{\dagger }}}$ is the electron creation
operator at site $i$, ${\bm{S}_{i}}$ the spin operator, and
${n_{i}}$ the number operator. The summation is over all the
nearest-neighbors, $\langle ij\rangle$. The Hilbert space is
constrained by the no-double-occupancy condition, i.e., $n_{i}\leq
1$. At half-filling, $n_{i}=1$, the system reduces to Mott
insulators (antiferromagnets) with a superexchange coupling, $J$.
Upon doping a hole into this system, $\sum_{i}n_{i}=N-1$ ($N$ the
number of the lattice sites), and the hopping process is triggered
as described by the first term of equation (\ref{a}), with $t$ the
hopping integral.

Below, we shall focus on bipartite lattices of $N=N_{x}\times
N_{y}$, where $N_{x}$ and $N_{y}$ are the numbers of sites along the
$x$ and $y$ directions, respectively. We shall study the ladders
with a finite $N_{y}$ (from $1$ to $5$) and sufficiently large
$N_{x}$. We set $J=1$ as the unit of energy and focus on the $t/J=3$
case unless otherwise specifically stated. For the numerical
simulation, we use fully open boundary conditions, and keep enough
number of states in each DMRG block. Excellent convergence is
achieved with total truncation error of the order of $\leq 10^{-7}$.

\begin{figure*}
\includegraphics[width=0.8\textwidth]{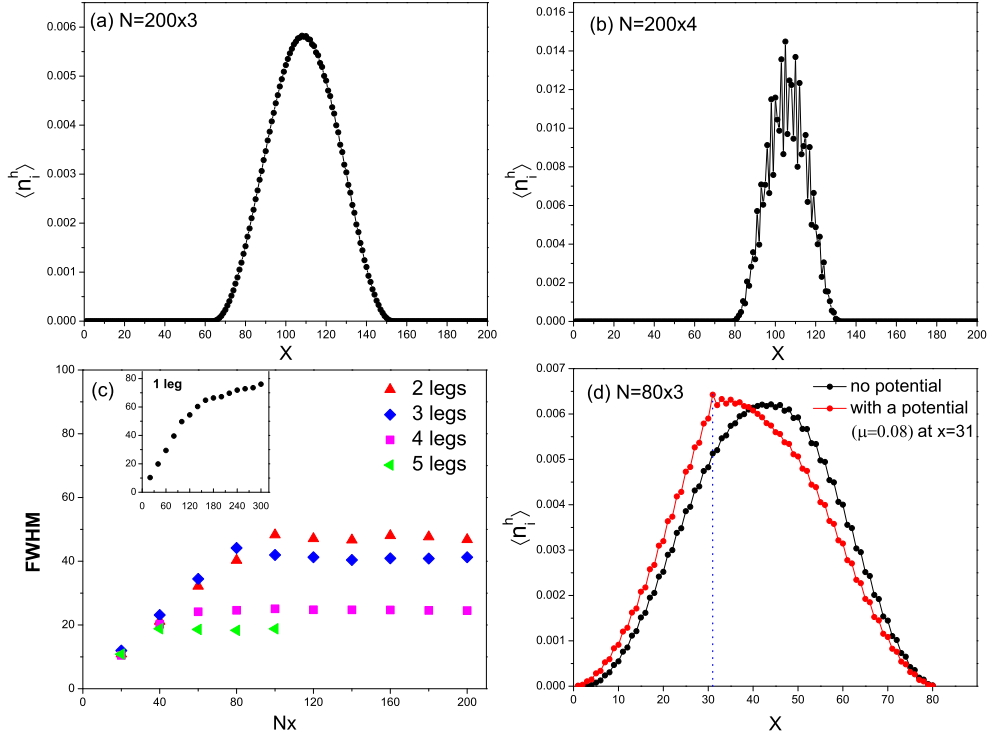} %
\renewcommand{\figurename}{Fig.}
\caption{\textbf{Self-localization of the doped hole.} In
the $3$-leg ladder system of size $N$=$%
200\times 3$ (a) and $4$-leg ladder of $N$=$200\times 4$ (b), single
hole doped into the antiferromagnets is well localized as shown by
the density distribution $\langle {n_{i}^{h}}\rangle $
in a middle leg (along the $x$ axis). $\langle
{n_{i}^{h}}\rangle $ is well fitted by the Gaussian function
with the full width at half-maximum (FWHM) defining the localization
length. The behavior of the FWHM depends on the width $N_{y}$ (c):
for the ladders of $N_{y}>1$, the FWHM first increases linearly at
small $N_{x}$ and then saturates at large $N_x$, with the saturation
values much smaller than the maximal $N_{x}$ ($=200$) and
monotonically decreasing with the leg number; for the
one-dimensional ($N_y=1$) chain, the FWHM increases monotonically
with the sample length without saturation (the inset of (c)). With
the open boundary condition, the hole is naturally localized at the
central region of the sample. Nevertheless, it can be easily
shuffled by adding a small local chemical potential $\mu$ as shown
in (d) (where $\mu=0.08$ at $x=31$).}\label{eg}
\end{figure*}

\emph{Self-localization of the doped hole.---} One of the main
findings of the present work is that the single hole is found to be
localized for the leg number $N_{y}>1$. Examples of the hole density
distribution, $\left \langle {n_{i}^{h}}\right \rangle \equiv 1-\left \langle {%
n_{i}}\right \rangle $, are shown in Fig.~\ref{eg} (a)
and (b) for $N_{y}=3$ and $4$, respectively, where the sample size
$N_{x}=200$ is clearly much larger than the localization lengths. In
Fig.~\ref{eg} (a) and (b), we plot the hole density at a middle leg
of the ladders along the $x$ direction. We have checked that upon
summing up the distribution from different legs, the sum rule
$\sum_{i}\left \langle {n_{i}^{h}}\right \rangle =1$
is satisfied. Examples of the contour plot of $%
\left \langle {n_{i}^{h}}\right \rangle $ in the $x$-$y$ plane can
be found in Supplementary Information.

To determine the localization length, we note that each density
profile is well fitted by a Gaussian distribution function. Then,
the localization length is defined as the full width at half-maximum
(FWHM) of the distribution. If the doped hole is itinerant, the FWHM
should increase monotonically with the sample length ($N_x$); in
contrast, if the doped hole is localized, the FWHM should saturate
at certain sample length, with the saturation value giving the
localization length. Fig.~\ref{eg} (c) shows that the FWHM for
each of the ladders, from $N_{y}=1$ to $5$, increases linearly for
small sample lengths, while saturates for sufficiently large sample
lengths (except the $N_{y}=1$ case, as shown in the inset of
Fig.~\ref{eg} (c)). These results indicate that the single hole is
well localized in space with a Gaussian profile of the density
distribution. In particular, Fig.~\ref{eg} (c) clearly shows that
the saturated FWHM at $N_{y}>1$ monotonically decreases with the
increase of the leg number $N_{y}$, implying the localization be
even stronger in the two-dimensional limit. By contrast, there is no
indication of saturation in FWHM for long one-dimensional ($N_y=1$)
chains with $N_{x}$ up to $300$ (see the inset of Fig.~\ref{eg}
(c)), which is consistent with the fact that the doped hole in one
dimensions is known to follow the Luttinger liquid behavior instead
of being localized\cite{Anderson,Weng1997}.

Figure~\ref{eg} (a) and (b) show that the hole is localized in the
central region of the sample. This is due to the fact that an open
boundary condition is used for the DMRG calculation. We also add an
artificial local chemical potential $\mu$($=0.08$) at the site
$x=31$, and find that the hole density profile is moved and peaks at
the impurity accordingly (see Fig.~\ref{eg} (d)). This suggests the
robustness of localization.

\begin{figure}
\begin{center}
\includegraphics[width=0.45\textwidth]{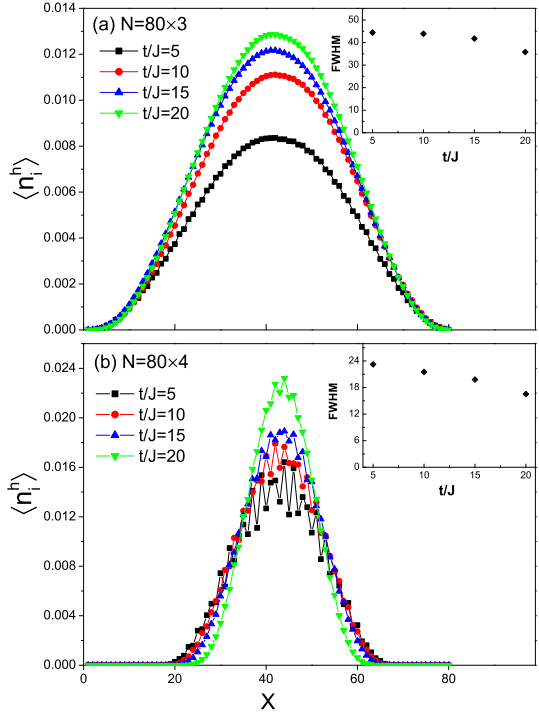}
\end{center}
\caption{\textbf{$t/J$-dependence of the hole density distribution.}
For different $t/J$ ratios, we plot $\langle n_{i}^{h}\rangle $ at
the middle chain along the $x$ axis for the ladders of size
$N=80\times 3$ (a) and $N=80\times 4$ (b). The localization length
(the FWHM) is reduced with the increase of $t/J$ (see the insets).
Moreover, the oscillation in the $4$-leg ladder system becomes
smaller with the increase of $t/J$, indicating the convergence of
the even- and odd-leg ladders.} \label{ratio-hole}
\end{figure}

Although the self-localization of the doped hole is insensitive to
the parity (even-odd) of the leg number, as shown in Fig.~\ref{eg}
(a) and (b) as well as in Supplementary Information, the hole
distribution function $\left \langle { n_{i}^{h}}\right \rangle $
does clearly exhibit a parity effect: for the even-leg ladders
($N_{y}=2,$ $4$), there are always small site-dependent oscillations
on top of the Gaussian density profile of $\left \langle
{n_{i}^{h}}\right \rangle $, while they are absent for the odd-leg
ladders ($N_{y}=3$, $5$). We point out that this can be attributed
to the underlying distinct spin-spin correlations for the odd- and
even-leg ladders at half-filling that follow a power-law (reflecting
the absence of spin gap) and exponential (reflecting the presence of
spin gap) decay, respectively (see Supplementary Information). In
fact, with the increase of the hopping integral $t$, the spin-gap
effect will be gradually reduced such that the parity effect
eventually disappears at large $t/J$ limit, where the aforementioned
oscillation in $\left \langle {n_{i}^{h}}\right \rangle $ for the
even-leg ladders is also diminished as illustrated in
Fig.~\ref{ratio-hole} (b). The localization length is actually
monotonically reduced as the ratio $t/J$ increases as shown in the
insets of Fig.~\ref{ratio-hole}. It thereby suggests that the
detailed spin dynamic behavior, governed by the superexchange $J$,
is not essential to the hole localization.

\emph{Vanishing quasiparticle spectral weight.---}One prominent
issue regarding the validity of quasiparticle picture concerns
whether the quasiparticle spectral weight $Z_{{\bm k}}$ vanishes at
the Fermi momentum. Despite of great efforts
\cite{Anderson90PRL,SCBA1,SCBA2,SCBA3,SCBA4,Dagotto1994,ED,Weng1996,Weng1997},
this remains highly controversial in the literature. In this part,
we turn to study this subject by invoking the state of the art
numerical method. The spectral weight $Z_{{\bm k}}$ is defined as
\begin{equation}
Z_{{\bm k}}\equiv \left | {\left \langle {\psi _{1-hole}}\right | c_{{\bm
k}\sigma }}\left | {\psi _{0-hole}}\right \rangle \right | ^{2},
\label{qpweight}
\end{equation}%
where $\left | {\psi _{1-hole}}\right \rangle $ is an eigenstate of
$H_{t-J}$ of momentum $-{\bm k}$ in the one hole case, $\left |{\psi
_{0-hole}}\right \rangle $ the ground state at half-filling, and
$c_{{\bm k}\sigma }$ is the Fourier transformation of the operator
$c_{i\sigma}$. Due to the translational symmetry of $H_{t-J}$,
$Z_{{\bm k}}$ is well-defined and characterizes the low-lying
quasiparticle-like excitations. A finite $ Z_{{\bm k}_{f}}$ means a
finite overlap of the bare-hole state $c_{{\bm k}_{f}\sigma }\left
|{\psi _{0-hole}}\right \rangle $ with the true ground state of a
single hole at the Fermi momentum $-{\bm k}_{f}$. If $Z_{{\bm
k}_{f}}=0$, then each injected hole will cause a global adjustment
in the ground state, rendering the breakdown of the quasiparticle
picture that is perturbatively tractable \cite{Anderson}.

\begin{figure}
\begin{center}
\includegraphics[width=0.45\textwidth]{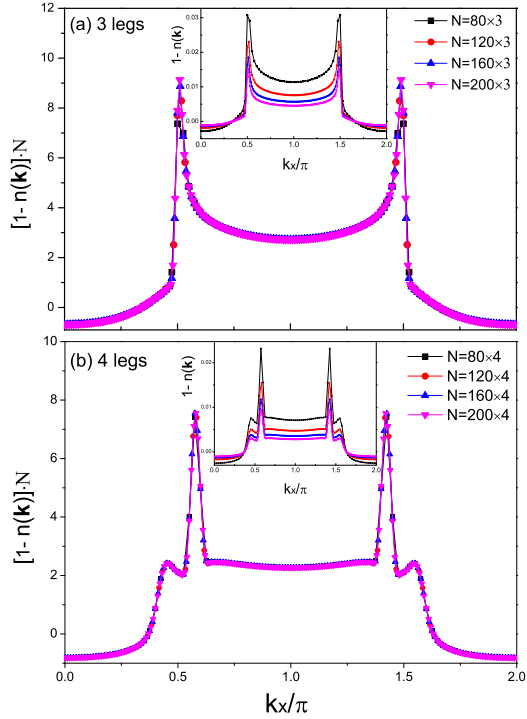}
\end{center}
\caption{\textbf{Quasiparticle spectral weight.} For both $3$-leg
(a) and $4$-leg (b) ladders, the momentum distribution of the hole
exhibits scaling behavior: after the rescaling, $1-n({\bm k})
\rightarrow [1-n({\bm k})]N$, the curves at different sample lengths
in the insets collapse into a single one represented in the
corresponding main panels. Note that for the $3$-leg case, we fix
$k_y=\frac{2\protect \pi}{3}$ and for the $4$-leg case,
$k_y=\frac{\protect \pi}{2}$ and plot the distribution along $k_x$.
Representative contour plots in the whole $k_x$-$k_y$ plane are shown
in Supplementary Information. } \label{collapse}
\end{figure}

Importantly, the finding of the self-localization of the single hole
implies that the doped hole should not behave like a conventional
quasiparticle. Consistent with this observation, the quasiparticle
spectral weight is numerically found to vanish, as we will show
below. Technically, different bases after truncation in the DMRG
calculation of $\left | {\psi _{0-hole}}\right \rangle $ and $\left
| {\psi _{1-hole}}\right \rangle $ make it difficult to directly
compute $Z_{{\bm k}_{f}}$ by using the definition (\ref{qpweight}).
To overcome this difficulty, we note that a finite $Z_{{\bm k}_{f}}$
implies a sudden jump in the momentum distribution function
$n(\bm{k})\equiv \sum_{\sigma } \langle {c_{\bm{k}\sigma
}^{\dag }c_{\bm{k}\sigma }}\rangle$ at the Fermi momentum.
Therefore, we first calculate $n(\bm{k})$, that is a task well
within the reach of the DMRG method, and then proceed to find
$Z_{{\bm k}_{f}}$. The results are shown in Fig.~\ref{collapse}.

The insets of Fig.~\ref{collapse} present the hole momentum
distribution $1-n(\bm{k})$ as a function of $k_{x}$ for fixed
$k_{y}=2\pi /3$ in the $3$-leg ladder (a) and $k_{y}=\pi /2$ in the
$4$-leg ladder (b). The value of $k_{y}$ is chosen in the way that
the ``sudden change'' in $1-n(\bm{k})$ can reach maxima by
varying $k_{x}$, according to the contour plots in the
$k_{x}$-$k_{y}$ plane (see Supplemental Information). It is very
important that after the rescaling: $1-n(\bm{k}) \rightarrow [
1-n(\bm{k})] N$, all the curves in the inset of
Fig.~\ref{collapse}(a) [or (b)] collapse into a universal curve
shown in the corresponding main panel. If one defines the Fermi
surface by the sudden jump in the momentum distribution function,
then the two universal curves suggest that the quasiparticle weight
$Z_{{\bm k}_{f}}$ scales as $1/N$ for large $N$, and vanishes in the
thermodynamic limit.

To satisfy the sum rule, $\sum_k [1-n(\bm k)]=1$, the width of the
jump in $1-n(\bm{k})$ must
remain finite in the limit $N\rightarrow \infty $. This is clearly shown in Fig. \ref%
{collapse}, consistent with a finite localization length in the real
space. Such a finite width in the momentum space may also look like
a remnant Fermi pocket of the hole (cf. the contour plots in
Supplementary Information), whose density is finite in the
localization volume even in the limit $N\rightarrow \infty $. We
have also calculated the momentum distribution of the hole at
different ratios of $t/J$. As shown in Supplementary Information,
for a given sample size, the jump near the Fermi point gets
continuously reduced with increasing $t/J$, which is consistent with
earlier work\cite{Dagotto1994}.

\begin{figure}
\begin{center}
\includegraphics[width=0.45\textwidth]{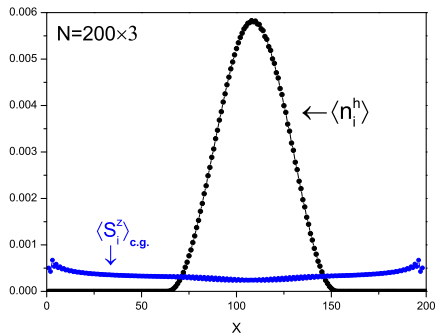}
\end{center}
\caption{\textbf{Spin-charge separation.} We remove a $\downarrow $
spin electron in a $N=200\times 3$ ladder and observe spin-charge
separation. The hole (charge) is localized at the sample center,
with the density distribution of Gaussian type; the spin of
$S^z=1/2$ spreads over the entire sample, with the (coarse grained)
density $\langle S^z_i\rangle_{c.g.}$ approximately uniform. Note
that to compute the spin density, a local coarse graining has been
performed so as to average out local antiferromagnetic oscillations.
} \label{scseparation}
\end{figure}

Moreover, it has been predicted analytically\cite{Weng2001} that a
spin-charge separation occurs as a spinless holon is localized while
a neutral spinon of $S=1/2$ moves away in the two-dimensional case.
The present numerical simulations indeed confirm this prediction for
the odd-leg ladders where the spin gap vanishes. Specifically, we
inject a hole into a $3$-leg ladder by removing a $\downarrow $ spin
electron. The results are shown in Fig.\ref{scseparation}: the hole
(charge) is localized at the sample center whereas the extra spin of
$S^z=1/2$ is spread along the $x$ direction, with a (coarse grained)
density $\left \langle S_{i}^{z}\right \rangle_{c.g.}$ approximately
uniform. (For the even-leg ladders, the spin gap renders the
observation of the spin-charge separation more difficult. Indeed, it
is required that the spin gap to be small enough, a condition that
can be achieved only if the leg number is sufficiently large.)

\emph{Discussion.---} We note that in Fig.~\ref{eg}, the
localization lengths are much larger than the lattice constant. This
indicates that the injected hole moves ``freely'' from one site to
the other within the localization volume. At larger (than the
localization length) scales, such a free motion is fully suppressed,
leading to the hole localization. In this sense, the scenario
resembles the quasi-one-dimensional Anderson localization discovered
by Efetov and Larkin long time ago\cite{Efetov83}. There, quantum
diffusion occurring at short scales is brought to a halt by wave
interference at large scales. The phenomenological analogy
notwithstanding, the two systems are fundamentally different: in the
present system, there are no extrinsic disorders at all. Instead, as
we will show below, the self-localization of the injected hole in
the quantum antiferromagnetic background arises from intrinsic
disorders of pure strong correlation origin. Therefore, it is
distinctly different from either conventional Anderson
localization\cite{Anderson58} or many-body
localization\cite{Basko06} discovered more recently.

\begin{figure}
\begin{center}
\includegraphics[width=0.45\textwidth]{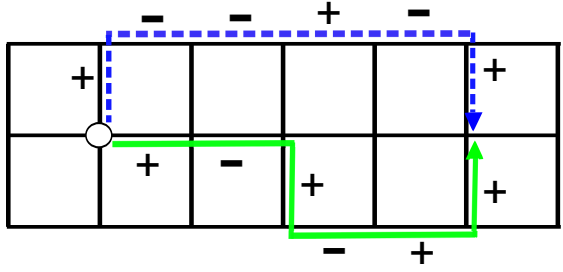}
\end{center}
\caption{\textbf{Interference picture of self-localization.} A sign
$\pm$ is acquired when the hole exchanges its position with the
nearest neighbor, an $\uparrow$ ($\downarrow$) spin. Consequently,
as the hole moves from the injection point to a distant site, a sign
sequence -- the phase string -- is left behind as exemplified by the
green (solid) and dashed (blue) lines. These phase strings
destructively interfere with each other, suppressing the forward
scattering and leading to the self-localization of the injected hole
in the quantum antiferromagnetic spin background.} \label{phase-int}
\end{figure}

Let us consider the motion of the hole in the quantum
antiferromagnetic spin background from the injection point to a
distant site (Fig.~\ref{phase-int}). The wave nature allows the hole
to propagate along all the possible paths. As discovered in
Refs.~\onlinecite{Weng1996,Weng1997}, given a hole path, $p$, the quantum
phase is completely determined by the parity of the
hole-$\downarrow$ spin exchange number, $N_h^\downarrow[p]$. More
precisely, upon hopping to its nearest neighbor, the hole acquires a
sign, $\pm $, depending on whether it exchanges its position with
$\uparrow$ or $\downarrow$ spin. As the hole moves along a given
path, a sign sequence known as the phase
string\cite{Weng1996,Weng1997} is left behind. The quantum amplitude
characterizing the hole motion is the superposition of all the
propagation amplitudes associated with individual hole paths, and is
given by
\begin{equation}\label{eq:1}
    \sum_{p}\sum_{\{\phi\}} \rho_p[\{\phi\}] (-1)^{N_h^\downarrow[p]},
\end{equation}
where the first sum is over all the possible hole paths, $p$, the
second over all the intermediate spin configurations, $\{\phi\}$,
encountered in the path, $p$, and $\rho_p[\{\phi\}]>0$. Equation
(\ref{eq:1}) suggests that as a unique property of Mott physics, the
spin configurations provide a translationally variant background --
the analog of disorder potentials -- for the hole motion.

To better understand the ``disordering'' nature of the phase
strings, we note that as the hole path is slightly distorted, the
intermediate spin configurations encountered may also be slightly
changed. But the resulting change in the {\it parity} of the number
$N_h^\downarrow$ can be significant. That is, the ``non-integrable
quantum phase factor'', $(-1)^{N_h^\downarrow[p]}$, will strongly
fluctuate as the hole paths are sampled. Then, from equation
(\ref{eq:1}), we see that even paths close to each other can
destructively interfere very singularly. As a result, the forward
scattering is significantly suppressed, and eventually, localization
is reached when the system is sufficiently large. Indeed, as the leg
number or the parameter $t/J$ increases, such destructive
interference proliferates, resulting in the decrease of the
localization length (Fig.~\ref{eg} and~\ref{ratio-hole}). In
contrast, provided that the leg number is unity, the destructive
interference from different paths no longer exists and the hole
escapes from localization, in agreement with the numerical
simulation (inset of Fig.~\ref{eg}). For the leg number comparable
to the sample length, the destructive interference arising from the
disordered signs leads to two-dimensional self-localization, as
proved analytically in Ref.~\onlinecite{Weng2001}. Therefore, we
conclude that the intrinsic ``disordered'' sign structure,
$(-1)^{N_h^\downarrow[p]}$, is responsible for the self-localization
of the injected hole and the vanishing quasiparticle spectral
weight.

In summary, we provide the first unambiguous numerical evidence
showing that a single hole injected into Mott antiferromagnets is
localized in the quantum antiferromagnetic background, and that the
quasiparticle weight vanishes in the thermodynamic limit. These
findings are consistent with both the analytic prediction of the
phase string theory for a single hole-doped two-dimensional Mott
antiferromagnet and the experimental results achieved by using ARPES
and STM methods. In view of that the phase string effect is a
rigorous property of the $t$-$J$ model on bipartite lattices
regardless of doping concentration, temperature, and
dimensions\cite{Wu-Weng-Zaanen}, the present work may provide
significant new insights into the long-standing issue of doped Mott
insulators and thereby high temperature superconductivity.

This work was supported by the NNSFC (under grant nos. 10834003 and
11174174), by the KITP (NSF no. PHY05-51164), by the NSF MRSEC Program
under Award (no. DMR 1121053), by the NBRPC (nos. 2009CB929402,
2010CB923003, 2011CBA00302 and 2011CBA00108), and by the Tsinghua
University Initiative Scientific Research Program (no. 2011Z02151).

\bibliography{Onehole}

\begin{thebibliography}{26}
\expandafter\ifx\csname natexlab\endcsname\relax\def\natexlab#1{#1}\fi
\expandafter\ifx\csname bibnamefont\endcsname\relax
  \def\bibnamefont#1{#1}\fi
\expandafter\ifx\csname bibfnamefont\endcsname\relax
  \def\bibfnamefont#1{#1}\fi
\expandafter\ifx\csname citenamefont\endcsname\relax
  \def\citenamefont#1{#1}\fi
\expandafter\ifx\csname url\endcsname\relax
  \def\url#1{\texttt{#1}}\fi
\expandafter\ifx\csname urlprefix\endcsname\relax\def\urlprefix{URL }\fi
\providecommand{\bibinfo}[2]{#2}
\providecommand{\eprint}[2][]{\url{#2}}

\bibitem[{\citenamefont{Anderson}(1997)}]{Anderson}
\bibinfo{author}{\bibfnamefont{P.~W.} \bibnamefont{Anderson}},
  \emph{\bibinfo{title}{{The Theory of Superconductivity in the High-Tc Cuprate
  Superconductors}}} (\bibinfo{publisher}{Princeton University Press,
  Princeton, NJ}, \bibinfo{year}{1997}).

\bibitem[{\citenamefont{Anderson}(1990)}]{Anderson90PRL}
\bibinfo{author}{\bibfnamefont{P.~W.} \bibnamefont{Anderson}},
  \bibinfo{journal}{Physical Review Letters} \textbf{\bibinfo{volume}{64}},
  \bibinfo{pages}{1839} (\bibinfo{year}{1990}).

\bibitem[{\citenamefont{Shraiman and Siggia}(1988)}]{Shraiman1988}
\bibinfo{author}{\bibfnamefont{B.~I.} \bibnamefont{Shraiman}} \bibnamefont{and}
  \bibinfo{author}{\bibfnamefont{E.~D.} \bibnamefont{Siggia}},
  \bibinfo{journal}{Physical Review Letters} \textbf{\bibinfo{volume}{61}},
  \bibinfo{pages}{467} (\bibinfo{year}{1988}).

\bibitem[{\citenamefont{Schmitt-Rink et~al.}(1988)\citenamefont{Schmitt-Rink,
  Varma, and Ruckenstein}}]{SCBA1}
\bibinfo{author}{\bibfnamefont{S.}~\bibnamefont{Schmitt-Rink}},
  \bibinfo{author}{\bibfnamefont{C.~M.} \bibnamefont{Varma}}, \bibnamefont{and}
  \bibinfo{author}{\bibfnamefont{A.~E.} \bibnamefont{Ruckenstein}},
  \bibinfo{journal}{Physical Review Letters} \textbf{\bibinfo{volume}{60}},
  \bibinfo{pages}{2793} (\bibinfo{year}{1988}).

\bibitem[{\citenamefont{Kane et~al.}(1989)\citenamefont{Kane, Lee, and
  Read}}]{SCBA2}
\bibinfo{author}{\bibfnamefont{C.~L.} \bibnamefont{Kane}},
  \bibinfo{author}{\bibfnamefont{P.~A.} \bibnamefont{Lee}}, \bibnamefont{and}
  \bibinfo{author}{\bibfnamefont{N.}~\bibnamefont{Read}},
  \bibinfo{journal}{Physical Review B} \textbf{\bibinfo{volume}{39}},
  \bibinfo{pages}{6880} (\bibinfo{year}{1989}).

\bibitem[{\citenamefont{Martinez and Horsch}(1991)}]{SCBA3}
\bibinfo{author}{\bibfnamefont{G.}~\bibnamefont{Martinez}} \bibnamefont{and}
  \bibinfo{author}{\bibfnamefont{P.}~\bibnamefont{Horsch}},
  \bibinfo{journal}{Physical Review B} \textbf{\bibinfo{volume}{44}},
  \bibinfo{pages}{317} (\bibinfo{year}{1991}).

\bibitem[{\citenamefont{Liu and Manousakis}(1991)}]{SCBA4}
\bibinfo{author}{\bibfnamefont{Z.}~\bibnamefont{Liu}} \bibnamefont{and}
  \bibinfo{author}{\bibfnamefont{E.}~\bibnamefont{Manousakis}},
  \bibinfo{journal}{Physical Review B} \textbf{\bibinfo{volume}{44}},
  \bibinfo{pages}{2414} (\bibinfo{year}{1991}).

\bibitem[{\citenamefont{Dagotto}(1994)}]{Dagotto1994}
\bibinfo{author}{\bibfnamefont{E.}~\bibnamefont{Dagotto}},
  \bibinfo{journal}{Reviews of Modern Physics} \textbf{\bibinfo{volume}{66}},
  \bibinfo{pages}{763} (\bibinfo{year}{1994}).

\bibitem[{\citenamefont{Leung and Gooding}(1995)}]{ED}
\bibinfo{author}{\bibfnamefont{P.~W.} \bibnamefont{Leung}} \bibnamefont{and}
  \bibinfo{author}{\bibfnamefont{R.~J.} \bibnamefont{Gooding}},
  \bibinfo{journal}{Physical Review B} \textbf{\bibinfo{volume}{52}},
  \bibinfo{pages}{R15711} (\bibinfo{year}{1995}).

\bibitem[{\citenamefont{Lee and Shih}(1997)}]{Shih97}
\bibinfo{author}{\bibfnamefont{T.~K.} \bibnamefont{Lee}} \bibnamefont{and}
  \bibinfo{author}{\bibfnamefont{C.~T.} \bibnamefont{Shih}},
  \bibinfo{journal}{Physical Review B} \textbf{\bibinfo{volume}{55}},
  \bibinfo{pages}{5983} (\bibinfo{year}{1997}).

\bibitem[{\citenamefont{White and Scalapino}(1997)}]{White2}
\bibinfo{author}{\bibfnamefont{S.~R.} \bibnamefont{White}} \bibnamefont{and}
  \bibinfo{author}{\bibfnamefont{D.~J.} \bibnamefont{Scalapino}},
  \bibinfo{journal}{Physical Review B} \textbf{\bibinfo{volume}{55}},
  \bibinfo{pages}{6504} (\bibinfo{year}{1997}).

\bibitem[{\citenamefont{Laughlin}(1997)}]{Laughlin97}
\bibinfo{author}{\bibfnamefont{R.~B.} \bibnamefont{Laughlin}},
  \bibinfo{journal}{Physical Review Letters} \textbf{\bibinfo{volume}{79}},
  \bibinfo{pages}{1726} (\bibinfo{year}{1997}).

\bibitem[{\citenamefont{Sheng et~al.}(1996)\citenamefont{Sheng, Chen, and
  Weng}}]{Weng1996}
\bibinfo{author}{\bibfnamefont{D.~N.} \bibnamefont{Sheng}},
  \bibinfo{author}{\bibfnamefont{Y.~C.} \bibnamefont{Chen}}, \bibnamefont{and}
  \bibinfo{author}{\bibfnamefont{Z.~Y.} \bibnamefont{Weng}},
  \bibinfo{journal}{Physical Review Letters} \textbf{\bibinfo{volume}{77}},
  \bibinfo{pages}{5102} (\bibinfo{year}{1996}).

\bibitem[{\citenamefont{Weng et~al.}(1997)\citenamefont{Weng, Sheng, Chen, and
  Ting}}]{Weng1997}
\bibinfo{author}{\bibfnamefont{Z.~Y.} \bibnamefont{Weng}},
  \bibinfo{author}{\bibfnamefont{D.~N.} \bibnamefont{Sheng}},
  \bibinfo{author}{\bibfnamefont{Y.-C.} \bibnamefont{Chen}}, \bibnamefont{and}
  \bibinfo{author}{\bibfnamefont{C.~S.} \bibnamefont{Ting}},
  \bibinfo{journal}{Physical Review B} \textbf{\bibinfo{volume}{55}},
  \bibinfo{pages}{3894} (\bibinfo{year}{1997}).

\bibitem[{\citenamefont{Weng et~al.}(2001)\citenamefont{Weng, Muthukumar,
  Sheng, and Ting}}]{Weng2001}
\bibinfo{author}{\bibfnamefont{Z.~Y.} \bibnamefont{Weng}},
  \bibinfo{author}{\bibfnamefont{V.~N.} \bibnamefont{Muthukumar}},
  \bibinfo{author}{\bibfnamefont{D.~N.} \bibnamefont{Sheng}}, \bibnamefont{and}
  \bibinfo{author}{\bibfnamefont{C.~S.} \bibnamefont{Ting}},
  \bibinfo{journal}{Physical Review B} \textbf{\bibinfo{volume}{63}},
  \bibinfo{pages}{075102} (\bibinfo{year}{2001}).

\bibitem[{\citenamefont{Ye et~al.}(2012)\citenamefont{Ye, Cai, Yu, Zhou, Ruan,
  Liu, Jin, and Wang}}]{STM}
\bibinfo{author}{\bibfnamefont{C.}~\bibnamefont{Ye}},
  \bibinfo{author}{\bibfnamefont{P.}~\bibnamefont{Cai}},
  \bibinfo{author}{\bibfnamefont{R.}~\bibnamefont{Yu}},
  \bibinfo{author}{\bibfnamefont{X.}~\bibnamefont{Zhou}},
  \bibinfo{author}{\bibfnamefont{W.}~\bibnamefont{Ruan}},
  \bibinfo{author}{\bibfnamefont{Q.}~\bibnamefont{Liu}},
  \bibinfo{author}{\bibfnamefont{C.}~\bibnamefont{Jin}}, \bibnamefont{and}
  \bibinfo{author}{\bibfnamefont{Y.}~\bibnamefont{Wang}}
  (\bibinfo{year}{2012}), \eprint{1201.0342}.

\bibitem[{\citenamefont{Anderson}(1958)}]{Anderson58}
\bibinfo{author}{\bibfnamefont{P.~W.} \bibnamefont{Anderson}},
  \bibinfo{journal}{Physical Review} \textbf{\bibinfo{volume}{109}},
  \bibinfo{pages}{1492} (\bibinfo{year}{1958}).

\bibitem[{\citenamefont{Basko et~al.}(2006)\citenamefont{Basko, Aleiner, and
  Altshuler}}]{Basko06}
\bibinfo{author}{\bibfnamefont{D.~M.} \bibnamefont{Basko}},
  \bibinfo{author}{\bibfnamefont{I.~L.} \bibnamefont{Aleiner}},
  \bibnamefont{and} \bibinfo{author}{\bibfnamefont{B.~L.}
  \bibnamefont{Altshuler}}, \bibinfo{journal}{Annals of Physics}
  \textbf{\bibinfo{volume}{321}}, \bibinfo{pages}{1126} (\bibinfo{year}{2006}).

\bibitem[{\citenamefont{Anderson}(1987)}]{anderson87}
\bibinfo{author}{\bibfnamefont{P.~W.} \bibnamefont{Anderson}},
  \bibinfo{journal}{Science (New York, N.Y.)} \textbf{\bibinfo{volume}{235}},
  \bibinfo{pages}{1196} (\bibinfo{year}{1987}).

\bibitem[{\citenamefont{Lee et~al.}(2006)\citenamefont{Lee, Nagaosa, and
  Wen}}]{Lee06}
\bibinfo{author}{\bibfnamefont{P.~A.} \bibnamefont{Lee}},
  \bibinfo{author}{\bibfnamefont{N.}~\bibnamefont{Nagaosa}}, \bibnamefont{and}
  \bibinfo{author}{\bibfnamefont{X.-G.} \bibnamefont{Wen}},
  \bibinfo{journal}{Reviews of Modern Physics} \textbf{\bibinfo{volume}{78}},
  \bibinfo{pages}{17} (\bibinfo{year}{2006}).

\bibitem[{\citenamefont{Wells et~al.}(1995)\citenamefont{Wells, Shen, Matsuura,
  King, Kastner, Greven, and Birgeneau}}]{Shen95}
\bibinfo{author}{\bibfnamefont{B.~O.} \bibnamefont{Wells}},
  \bibinfo{author}{\bibfnamefont{Z.-X.} \bibnamefont{Shen}},
  \bibinfo{author}{\bibfnamefont{A.}~\bibnamefont{Matsuura}},
  \bibinfo{author}{\bibfnamefont{D.~M.} \bibnamefont{King}},
  \bibinfo{author}{\bibfnamefont{M.~A.} \bibnamefont{Kastner}},
  \bibinfo{author}{\bibfnamefont{M.}~\bibnamefont{Greven}}, \bibnamefont{and}
  \bibinfo{author}{\bibfnamefont{R.~J.} \bibnamefont{Birgeneau}},
  \bibinfo{journal}{Physical Review Letters} \textbf{\bibinfo{volume}{74}},
  \bibinfo{pages}{964} (\bibinfo{year}{1995}).

\bibitem[{\citenamefont{Ronning et~al.}(1998)\citenamefont{Ronning, Kim, Feng,
  Marshall, Loeser, Miller, Eckstein, Bozovic, and Shen}}]{ARPES}
\bibinfo{author}{\bibfnamefont{F.}~\bibnamefont{Ronning}},
  \bibinfo{author}{\bibfnamefont{C.}~\bibnamefont{Kim}},
  \bibinfo{author}{\bibfnamefont{D.~L.} \bibnamefont{Feng}},
  \bibinfo{author}{\bibfnamefont{D.~S.} \bibnamefont{Marshall}},
  \bibinfo{author}{\bibfnamefont{A.~G.} \bibnamefont{Loeser}},
  \bibinfo{author}{\bibfnamefont{L.~L.} \bibnamefont{Miller}},
  \bibinfo{author}{\bibfnamefont{J.}~\bibnamefont{Eckstein}},
  \bibinfo{author}{\bibfnamefont{I.}~\bibnamefont{Bozovic}}, \bibnamefont{and}
  \bibinfo{author}{\bibfnamefont{Z.-X.} \bibnamefont{Shen}},
  \bibinfo{journal}{Science} \textbf{\bibinfo{volume}{282}},
  \bibinfo{pages}{2067} (\bibinfo{year}{1998}).

\bibitem[{\citenamefont{Shen et~al.}(2004)\citenamefont{Shen, Ronning, Lu, Lee,
  Ingle, Meevasana, Baumberger, Damascelli, Armitage, Miller et~al.}}]{Shen04}
\bibinfo{author}{\bibfnamefont{K.~M.} \bibnamefont{Shen}},
  \bibinfo{author}{\bibfnamefont{F.}~\bibnamefont{Ronning}},
  \bibinfo{author}{\bibfnamefont{D.~H.} \bibnamefont{Lu}},
  \bibinfo{author}{\bibfnamefont{W.~S.} \bibnamefont{Lee}},
  \bibinfo{author}{\bibfnamefont{N.~J.~C.} \bibnamefont{Ingle}},
  \bibinfo{author}{\bibfnamefont{W.}~\bibnamefont{Meevasana}},
  \bibinfo{author}{\bibfnamefont{F.}~\bibnamefont{Baumberger}},
  \bibinfo{author}{\bibfnamefont{A.}~\bibnamefont{Damascelli}},
  \bibinfo{author}{\bibfnamefont{N.~P.} \bibnamefont{Armitage}},
  \bibinfo{author}{\bibfnamefont{L.~L.} \bibnamefont{Miller}},
  \bibnamefont{et~al.}, \bibinfo{journal}{Physical Review Letters}
  \textbf{\bibinfo{volume}{93}}, \bibinfo{pages}{267002}
  (\bibinfo{year}{2004}).

\bibitem[{\citenamefont{White}(1992)}]{White1992}
\bibinfo{author}{\bibfnamefont{S.~R.} \bibnamefont{White}},
  \bibinfo{journal}{Physical Review Letters} \textbf{\bibinfo{volume}{69}},
  \bibinfo{pages}{2863} (\bibinfo{year}{1992}).

\bibitem[{\citenamefont{Efetov and Larkin}(1983)}]{Efetov83}
\bibinfo{author}{\bibfnamefont{K.~B.} \bibnamefont{Efetov}} \bibnamefont{and}
  \bibinfo{author}{\bibfnamefont{A.~I.} \bibnamefont{Larkin}},
  \bibinfo{journal}{Sov. Phys. JETP} \textbf{\bibinfo{volume}{58}},
  \bibinfo{pages}{444} (\bibinfo{year}{1983}).

\bibitem[{\citenamefont{Wu et~al.}(2008)\citenamefont{Wu, Weng, and
  Zaanen}}]{Wu-Weng-Zaanen}
\bibinfo{author}{\bibfnamefont{K.}~\bibnamefont{Wu}},
  \bibinfo{author}{\bibfnamefont{Z.~Y.} \bibnamefont{Weng}}, \bibnamefont{and}
  \bibinfo{author}{\bibfnamefont{J.}~\bibnamefont{Zaanen}},
  \bibinfo{journal}{Physical Review B} \textbf{\bibinfo{volume}{77}},
  \bibinfo{pages}{155102} (\bibinfo{year}{2008}).

\end{thebibliography}


\begin{thebibliography}{9}
\bibitem{White94} S.R.White, R. M. Noack, and D. J. Scalapino. Phys. Rev.
Lett. \textbf{73}, 886(1994).
\end{thebibliography}

\newpage
\cleardoublepage

\begin{center}
{\large\bf SUPPLEMENTAL INFORMATION \\
Self-localization of a single hole in Mott antiferromagnets}
~\\
Zheng Zhu$^1$, Hong-Chen Jiang$^{2,3}$, Yang Qi$^1$,
Chu-Shun Tian$^1$ and Zheng-Yu Weng$^1$\\
{\em$^{1}$Institute for Advanced Study, Tsinghua University, Beijing, 100084,
China\\
$^2$Kavli Institute for Theoretical Physics, University of California,\\ Santa
Barbara, CA, 93106, U.S.A.\\
$^3$Center for Quantum Information, IIIS, Tsinghua University, Beijing,
100084, China}\\
{\small (Dated: \today)}\\
\end{center}

\renewcommand{\thefigure}{S\arabic{figure}} \renewcommand{\theequation}{S%
\arabic{equation}}
\renewcommand{\thesection}{\Roman{section}} \setcounter{section}{0} %
\setcounter{figure}{0}
\setcounter{equation}{0}





\begin{figure}[tbp]
\begin{center}
\includegraphics[width=10cm,height=6cm]{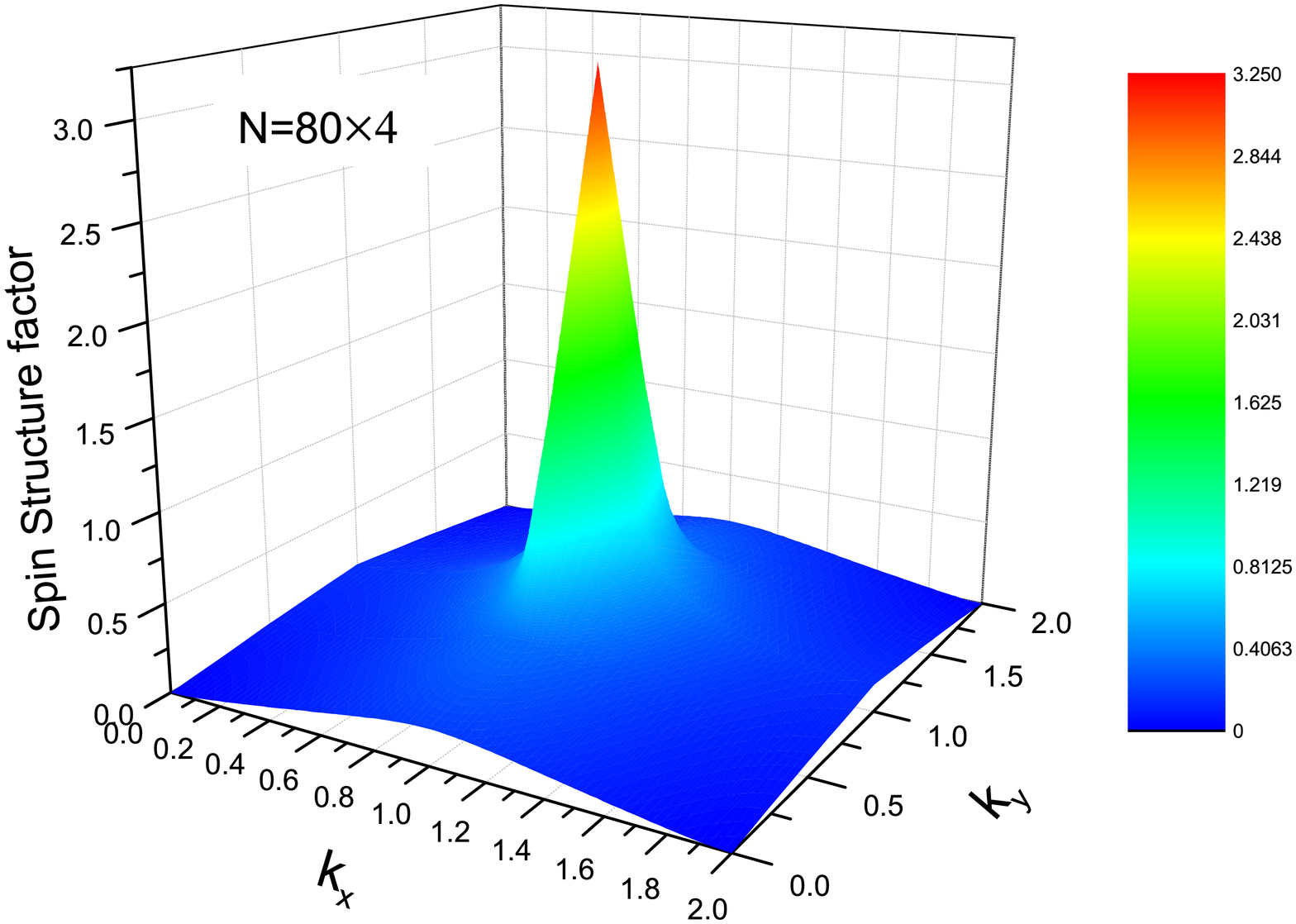}
\end{center}
\par
\renewcommand{\figurename}{Fig.}
\caption{Spin structure factor for the 4-leg ladder ($N$=80$\times4$), in
which the peak is located at ($\protect \pi$,$\protect \pi$) indicating strong
spin antiferromagnetic correlations even though a finite spin gap is
present. }
\label{structurefactor}
\end{figure}

\renewcommand{\theequation}{S\arabic{equation}}
\setcounter{equation}{0}
At half-filling, the $t$-$J$ model reduces to the Heisenberg model. For the
isotropic Heisenberg coupled-chain systems, the behavior of the even-leg
ladders is dramatically different from that of the odd ones. A well-known
fact is that the even-leg ladders have a spin gap, while the odd-leg ladders
are gapless, leading to an exponential decay of the spin-spin correlation in
the former, while a power law decay in the latter (see Fig.~\ref%
{spincorrelation}). The spin gap for the even-leg ladders is expected to
vanish in the large leg number (namely two-dimensional) limit. The spin
structure factor for the $4$-leg case has already shown strong
antiferromagnetic correlations as illustrated in Fig.~\ref{structurefactor}.
These results are consistent with earlier DMRG work\cite{White94}.

\begin{figure}[hbp]
\begin{center}
\includegraphics[width=0.9\textwidth]{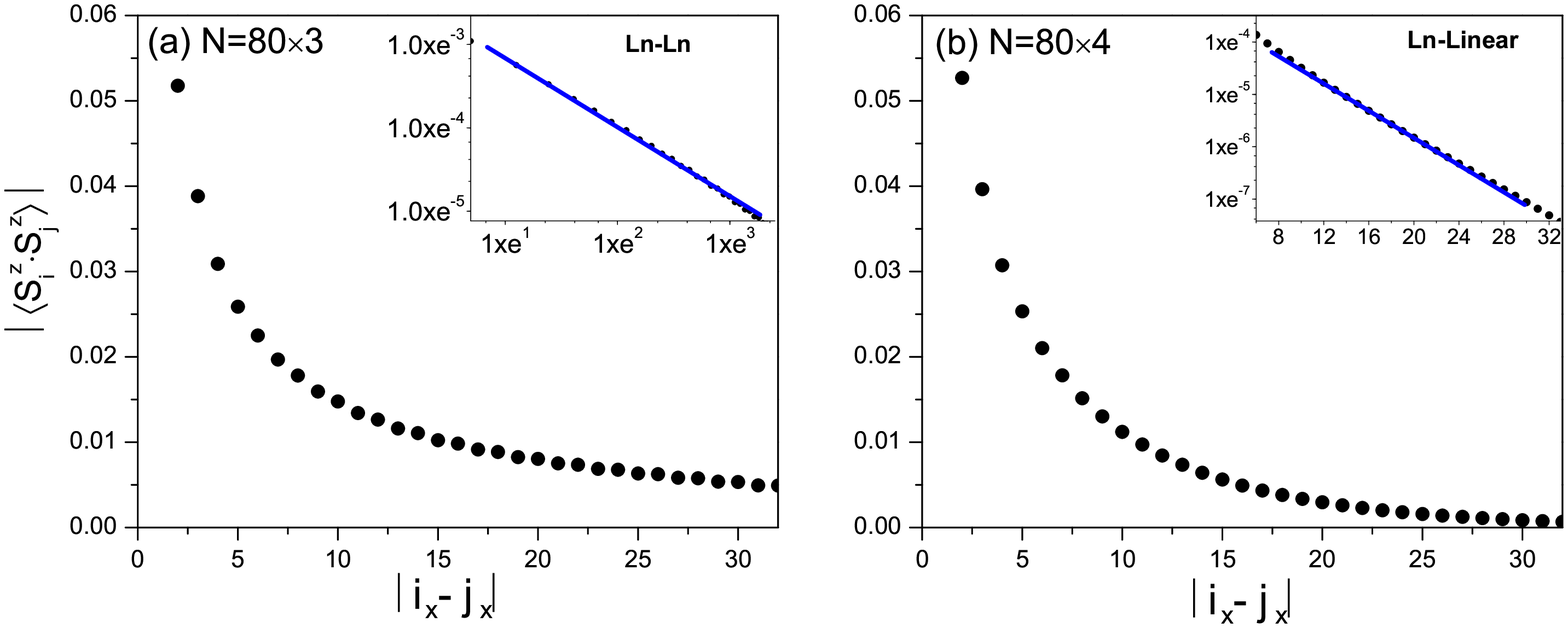}
\end{center}
\par
\renewcommand{\figurename}{Fig.}
\caption{The main panels show the spin-spin correlation $\left| {\left
\langle {S_i^z S_j^z } \right \rangle } \right|$ versus $\left| {i_x - j_x }
\right|$ with $i_x$ and $j_x$ located on the middle leg of $3$-leg (a) and
4-leg ladder (b), respectively. The insets are $\ln$-$\ln$ plot in (a) and $%
\ln$-linear plot (b), fitted with the straight (blue) lines which indicate
the power law decay for the former and exponential decay for the latter.}
\label{spincorrelation}
\end{figure}

\begin{figure}[tbp]
\begin{center}
\includegraphics[width=0.45\textwidth]{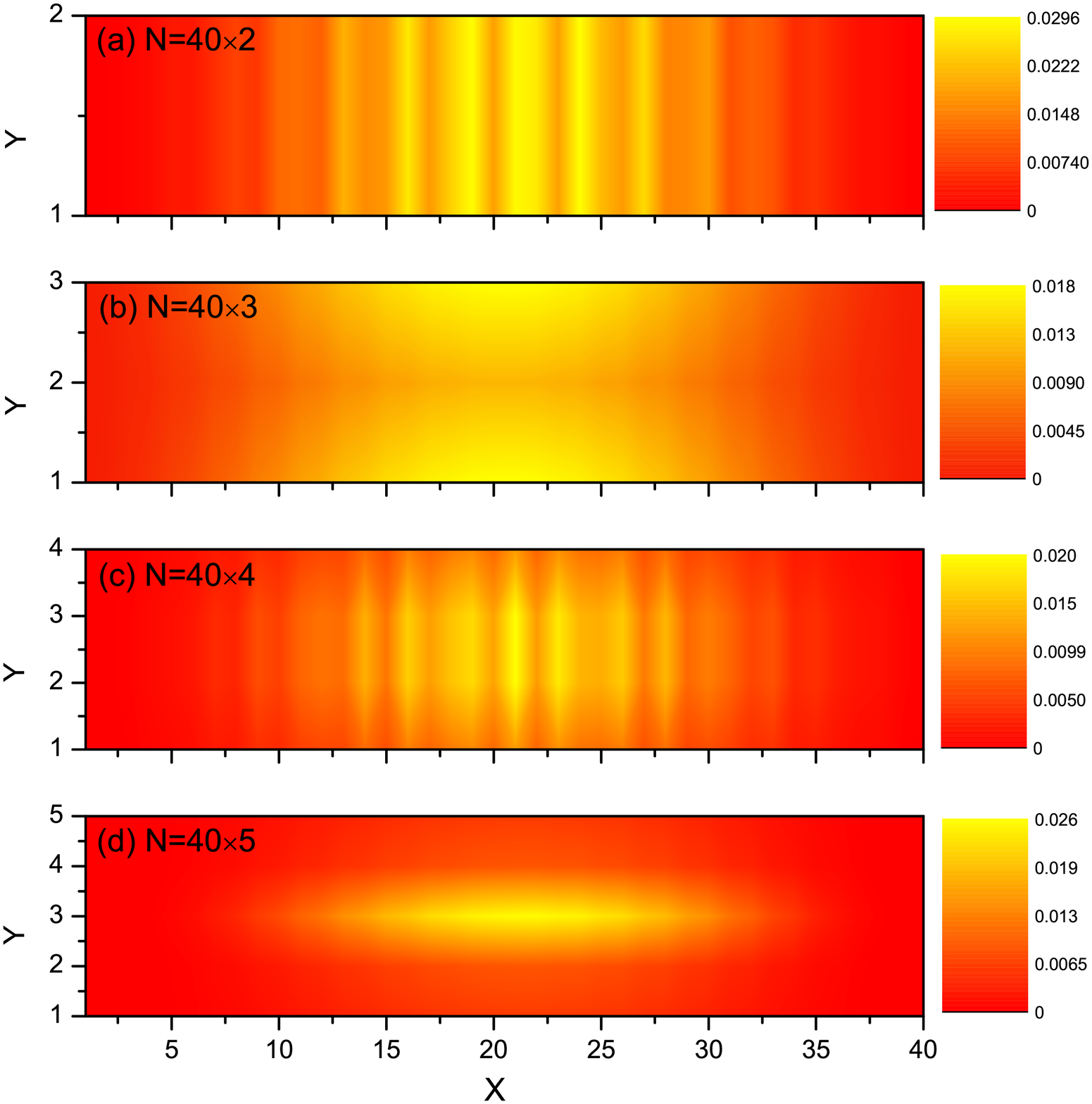}
\end{center}
\par
\renewcommand{\figurename}{Fig.}
\caption{Contour plots of the hole distribution in real space. The
sample size $N$ is $40\times 2$ (a), $40\times 3$ (b), $40\times 4$
(c), and $40\times 5$ (d) from top down.}  \label{Total_Contour}
\end{figure}

For the hole density distribution $\langle {n_{i}^{h}}\rangle $ calculated
in this work, the total hole number is fixed, $\sum \limits_{i}\langle {%
n_{i}^{h}}\rangle =1$. The contour plot $\langle {n_{i}^{h}}\rangle $ in the
$x$-$y$ plane of the ladder is illustrated in Fig.~\ref{Total_Contour} for $%
N=40\times N_{y}$ with $N_{y}=2$, $3$, $4$, and $5$. Besides the
localization in the central region along the chain ($x$) direction, a
prominent feature in the profiles is the distinction between the even- and
odd-leg ladders: the former has a spatial oscillation, while the latter has
none. This is concomitant with the parity effect in the spin-spin
correlation seen in Fig.~\ref{spincorrelation}.
\begin{figure}[tbp]
\centering
\subfigure[N=80$\times$3]{\includegraphics[width=0.45%
\textwidth]{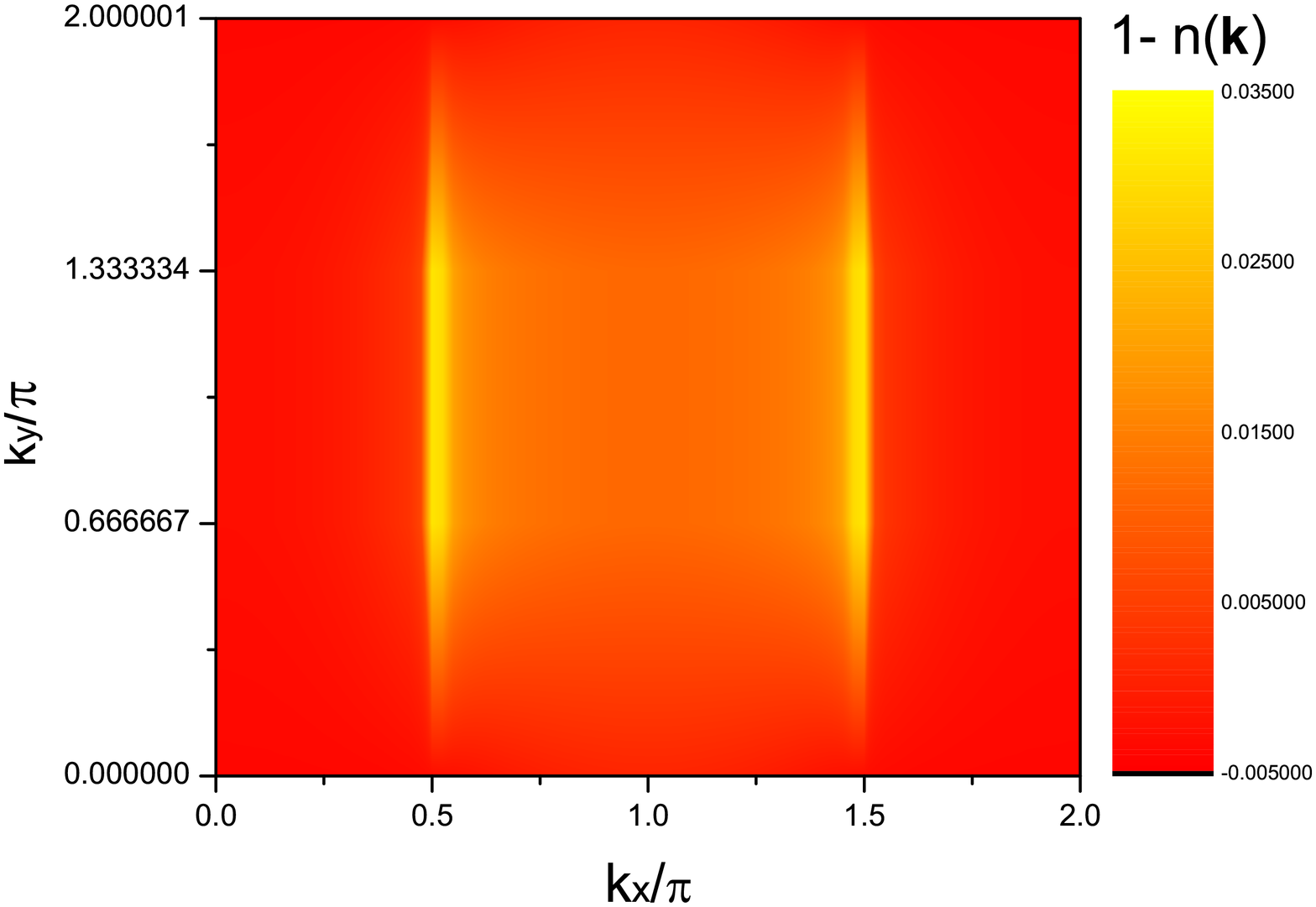}} 
\subfigure[N=80$\times$4]{\includegraphics[width=0.45%
\textwidth]{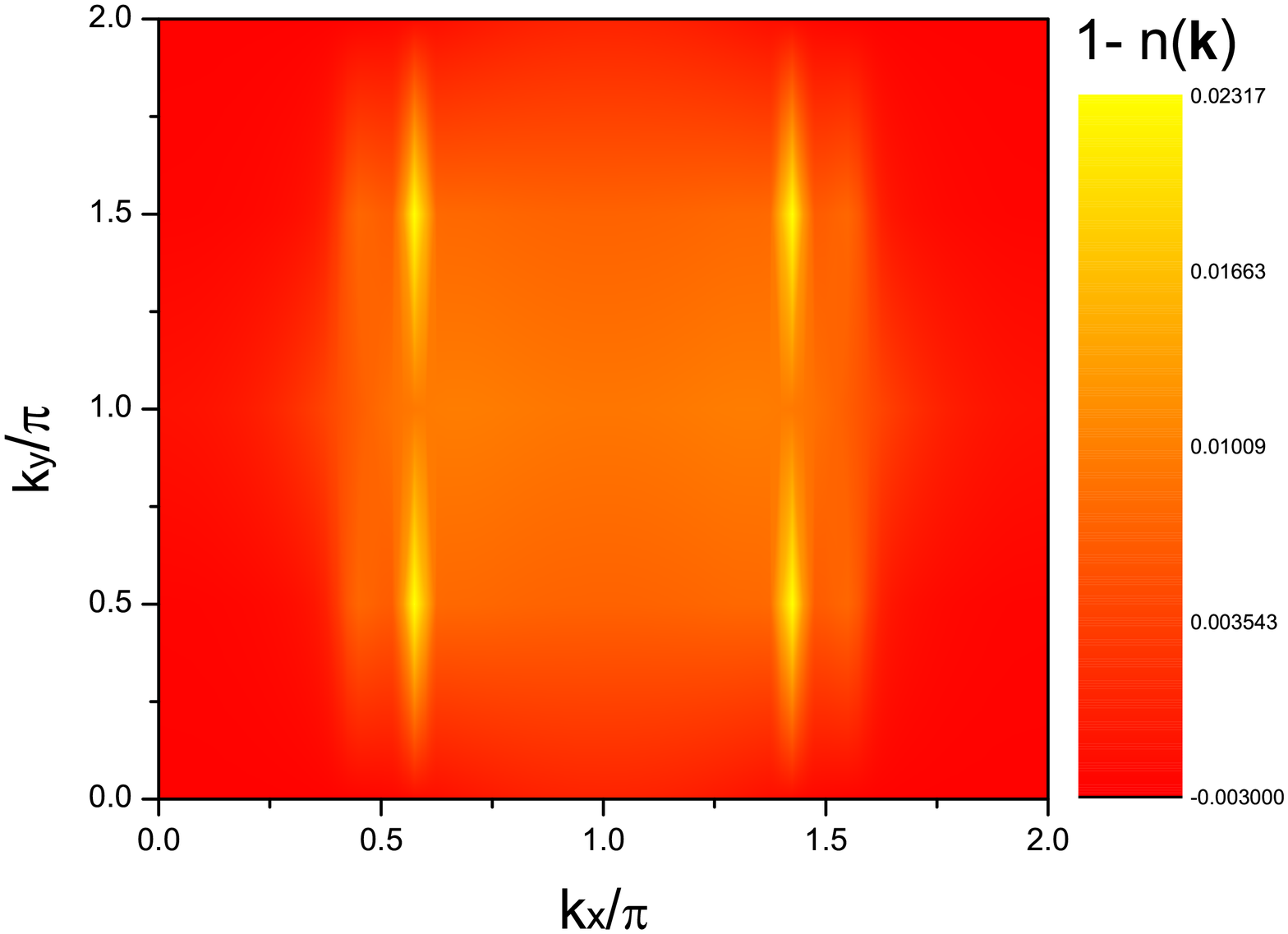}} \renewcommand{%
\figurename}{Fig.}
\caption{Contour plots of the electron momentum distribution $n(\mathbf{k})$
for $3$-leg (a) and $4$-leg (b) ladders. }
\label{n(k)contour}
\end{figure}
\begin{figure}[tbp]
\begin{center}
\includegraphics[width=0.9\textwidth]{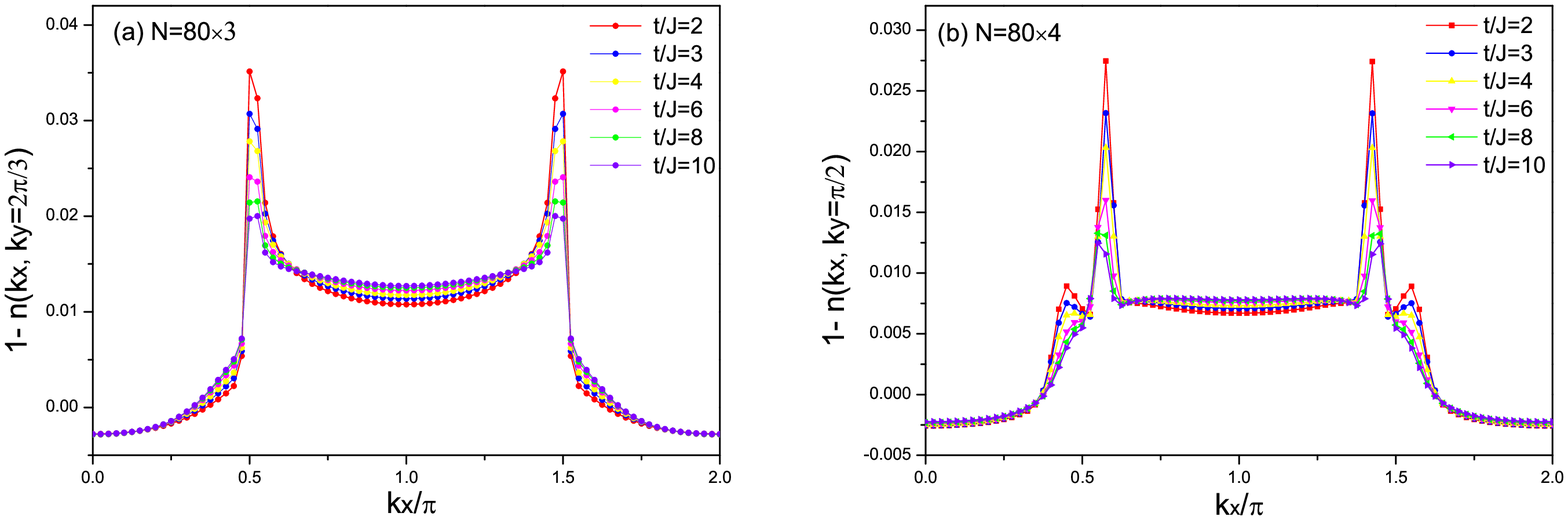}
\end{center}
\par
\renewcommand{\figurename}{Fig.}
\caption{The hole momentum distribution at different ratios of $t/J$ for $3$%
-leg (a) and $4$-leg (b) ladders. }
\label{ratio}
\end{figure}

Fig.~\ref{n(k)contour} shows the contour plot of the electron momentum
distribution function $n(\mathbf{k})$ for the $3$-leg (a) and $4$-leg (b)
cases, respectively at $N_x=80$. The minimum of $n(\mathbf{k})$ appears at $%
k_y$=$\pm \pi/2$ for the $4$-leg ladder and $k_y$=$\pm2\pi/3$ for the $3$%
-leg ladder. The jump of the hole momentum distribution as a
function of the ratio $t/J$ is shown in Fig.~\ref{ratio}, which
indicates that the jump is monotonically reduced with the increase
of $t/J$ .

\end{document}